\documentclass[aps,prl,showpacs,twocolumn,superscriptaddress]{revtex4}
\usepackage{amsfonts,amssymb,amsmath}
\usepackage{color}
\usepackage{graphicx}
\usepackage{dcolumn}
\usepackage{bm}

\begin{document}

\title{Demixing of compact polymers chains in three dimensions}

\author{Jesper Lykke Jacobsen}

\affiliation{LPTENS, \'Ecole Normale Sup\'erieure,
 24 rue Lhomond, 75231 Paris, France}
\affiliation{Universit\'e Pierre et Marie Curie,
 4 place Jussieu, 75252 Paris, France}

\date{July 8, 2010. Revised: October 29, 2010.}

\begin{abstract}

  We present a Monte Carlo algorithm that provides efficient and
  unbiased sampling of polymer melts consisting of two chains of equal
  length that jointly visit all the sites of a cubic lattice with rod
  geometry $L \times L \times rL$ and non-periodic (hard wall)
  boundary conditions. Using this algorithm for chains of length up to
  $40\,000$ monomers and aspect ratios $1 \le r \le 10$, we show that
  in the limit of a large lattice the two chains phase separate. This
  demixing phenomenon is present already for $r=1$, and becomes more
  pronounced, albeit not perfect, as $r$ is increased.

\end{abstract}


\maketitle

\section{Introduction}

When a single polymer chain with excluded volume interactions is
brought into a bad solvent, the solvent molecules strongly repel those
of the chain. To minimize the interface between solvent and chain
molecules, the chain will collapse and expel the solvent molecules
from its interior. Similarly, a few polymer chains in a very large
volume of solvent will form a dilute solution of collapsed chains. In
particular, the individual chains will not mix \cite{deGennes}.

Consider now squeezing the chains together so as to form a solution in
which the monomer density is finite (and the solvent density can be
taken to zero). We may then ask whether different chains will mix or
segregate. One would expect that the answer to this question might
depend on how the number of chains scales with the size of the system.

A well-studied case is that of a polymer melt where many chains are
present. If there are of the order $N^{1/2}$ chains, occupying a total
volume $N$ (i.e., with an average monomer density of $N^{-1/2}$ for
each chain) the so-called Flory theorem \cite{FloryThm} is supposed to
apply. According to this, the repulsive hard-core interaction between
different chains is completely screened out at large distances. This
implies that a single chain within the melt exhibits ideal, Gaussian
behavior: its radius of gyration scales as the square root of the
chain length. In particular, different chains should mix
(interpenetrate) completely, as opposed to a demixing (segregation)
scenario where each chain would occupy a small lump of the available
space \cite{deGennes}.

It has recently been shown by the Strasbourg group in a number of
studies \cite{Wittmer} that the Flory theorem is not quite correct. In
particular, these authors have shown that the interplay between chain
connectivity and the incompressibility of the melt leads to an
effective repulsion between chain segments. The bond-bond correlation
function, supposed to be short-range by the Flory argument, turns out
to exhibit a long-range algebraic decay. A number of deviations from
the ideal Flory behavior have been pointed out, and shown to be
accounted for in an improved scaling theory \cite{Wittmer} that views
the chain as a hierarchical arrangement of correlation holes
\cite{deGennes} of sub-chains on all length scales.

On the other hand, one might be interested in the situation where
there is only a {\em finite} number of chains in a solution of finite
monomer density.  In this context, the issue of polymer mixing has
recently gained renewed interest in the area of physical biology. One
outstanding question is to understand the organization and segregation
of chromosomes within simple bacteria, including during duplication.
In particular, the bacterium {\em Escherichia coli} with a single
circular chromosome in a rod-shaped cell has served as a model system.
In this situation the radius of gyration of a single chain will
necessarily scale as the cube root of the chain length, and there is
no good reason why the Flory theory \cite{FloryThm} and the more
recent improvements of it \cite{Wittmer} should apply. The question
whether the chains will mix in that case therefore needs to be settled
separately, as do the possible consequences for chromosome
organization.

The behavior of a dense solution of just two polymer chains confined to a
bar-shaped geometry, akin to that of a rod-shaped cell, was discussed
in \cite{Jun}. A simple piston model, combined with a blob-type
analysis \cite{deGennes} (see also \cite{BrochardGennes}),
was used to point out the mechanisms by
which demixing might occur. In the limit of large piston pressure
(corresponding to maximal density) the two chains were nevertheless
found to mix. In contrast with this, studies of unentangled ring
polymers in concentrated solutions \cite{Muller,JunMulder} concluded
that the existence of topological barriers would lead to segregation.
A molecular dynamics study, using a parameter-free bead-spring polymer
model of a ring polymer melt, corroborated the segregation scenario
and found several structural and dynamical results in agreement with
experiments on {\em Drosophila} and budding yeast chromosomes
\cite{RosaEveraers}. The results of \cite{RosaEveraers} imply that a
certain number of experimental observations could be due to generic
polymer effects, and as such could be accounted for in very simple
models.

In this paper, we revisit the problem of two open chains (not rings).
In line with the above reductionist point of view, we do so by
studying numerically a simple and very precisely defined
parameter-free lattice model of a two-chain melt of unit monomer
density. This model accounts fully for the constraints of maximal
monomer density, and the self and mutual avoidance of the chains.

The conformations of a single maximally compact (space filling) chain
can be modelled on a lattice as a Hamiltonian walk (HW). By
definition, a HW is a self-avoiding walk that visits each lattice site
exactly once.  Upon adding further local interactions, the HW model
has been proposed as a description of protein melting \cite{Flory}
(with bending rigidity), or of protein folding \cite{Dill} (with
suitable interactions among amino acids).  The simplest example of the
latter is the so-called HP model in which only two types of amino
acids (Hydrophobic or Polar) are taken into account.

To similarly model a melt of a finite number of chains, we can use the
model of Hamiltonian chains (HC) \cite{HamChains} in which $M$ self-
and mutually avoiding walks jointly visit all lattice sites. By
definition, each chain has length at least one (its two end points
cannot coincide). In what follows we shall always take the underlying
lattice to be simple cubic (or square, when we occasionally compare
with the two-dimensional case). It seems natural to assume that the HC
model correctly describes an $M$-chain polymer melt in the large-scale limit,
where lattice details should be irrelevant.

The question is then whether the HC model validates, for the case of a
melt consisting of only a few chains, the mixing scenario envisaged
for the two-chain case in \cite{Jun}. In this paper we study in
detail the simplest possible case of $M=2$ identical chains of equal
length, by large-scale numerical simulations. We show that, contrary
to the above expectations, the chains actually phase separate. For an
$L \times L \times (rL)$ lattice, with large $L$ and fixed aspect
ratio $r$, the effect turns out to be quite subtle (but measurable)
for $r=1$, becoming more pronounced for larger values of $r$.

\section{Algorithm}

Properties of self-avoiding walks are routinely studied by importance
sampling schemes of the Monte Carlo (MC) type, one well-known example
being the pivot algorithm \cite{pivot}. Unfortunately, almost all
known algorithms become useless in the fully-packed limit, as the
acceptance ratio for any proposed move tends to zero. For the HW case
(i.e., the HC model with $M=1$ chain) we have recently shown that an
MC move proposed by Mansfield \cite{Mansfield} can be turned into an
algorithm that is both efficient (with dynamical exponent $z \simeq
0$) and provides unbiased sampling (each conformation is generated
with uniform probability) \cite{Backbite}.

\begin{figure}
\includegraphics[width=7cm,angle=0]{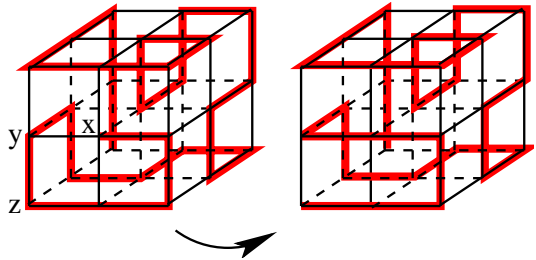}
\caption{(Color online.) Move between two of the $2\,480\,304$ possible Hamiltonian walks
on a $3 \times 3 \times 3$ cubic lattice.}
\label{fig:move}
\end{figure}

The working principle of this $M=1$ algorithm is shown in
Fig.~\ref{fig:move}. Choose an end point ${\bf x}$ of the HW at
random, and select one of its adjacent empty links $({\bf x}{\bf y})$.
Then, among the occupied links adjacent to ${\bf y}$, there is exactly
one which would form part of a loop if $({\bf x}{\bf y})$ were added
to the HW; call it $({\bf y}{\bf z})$. The basic move is then to
change the HW by adding $({\bf x}{\bf y})$ and removing $({\bf y}{\bf
  z})$.

We now state the modifications necessary to deal with the case of
$M=2$.  For the moment we assume that the point ${\bf x}$ is in the
bulk of the system; the case when it is on the boundary will be discussed
later.

Let the configuration ${\cal C}$ consist of two chains, and let ${\bf
  x}$ be one of the four end points chosen at random. Once again,
choose with uniform probability one of its adjacent empty links $({\bf
  x}{\bf y})$. As before, the idea will be to modify ${\cal C}$ by
adding $({\bf x}{\bf y})$ and removing one other link. But before
adding the link $({\bf x}{\bf y})$, the point ${\bf y}$ may be
adjacent on one or two occupied links. The former happens if ${\bf y}$
is another end point, either of the same chain as ${\bf x}$ or of the
other chain; in that case the choice of the link $({\bf y}{\bf z})$ to
be removed is unambiguous. In the latter case (i.e., when ${\bf y}$ is
adjacent on two occupied links) one must distinguish two different
situations.  If ${\bf y}$ is on the same chain as ${\bf x}$, one
proceeds as in the $M=1$ case, making the unique choice of ${\bf z}$
that avoids the formation of a loop. And if ${\bf x}$ and ${\bf y}$
are on different chains, one choses randomly between the two
possibilities for ${\bf z}$.

It remains to discuss two subtleties. First, when ${\bf x}$ and ${\bf
  y}$ are end points of two different chains, the move will consist in
letting the first (resp.\ second) chain grow (resp.\ retract) by one
monomer.  In particular, it might happen that ${\bf z}$ is the {\em
  other} end point of the second chain. The move would then consist in
letting the second chain shrink to zero, but since this is not an
allowed HC configuration, we remedy the situation by leaving ${\cal
  C}$ unchanged. Second, when ${\bf x}$ is at the boundary of the
system, it might happen that ${\bf y}$ is chosen outside the
lattice. In that case as well the move is rejected, i.e., ${\cal C}$
is left unchanged.

\section{Unbiased sampling} 

With all these rules, and considering that an MC move takes unit time
regardless of whether it leads to a change in ${\cal C}$ or not, it
follows from a careful analysis that the rule of detailed balance is
satisfied.  As an independent numerical check, we ran the simulation
on a $2 \times 2 \times 3$ cubic lattice until each of the $4\,204$
possible configurations \cite{HamChains} had been generated $\simeq 5
\cdot 10^4$ times. We then verified that the fluctuations in the
number of occurencies of each configuration was compatible with
counting statistics.

Proving that the MC move is also ergodic, i.e., that any desired
configuration may be reached from an initial one in a finite number of
steps, unfortunately turns out to be much harder (a proof does not
even exist for the $M=1$ chain case \cite{Backbite}).

As in \cite{Backbite}, we therefore resort to explicit checks of
ergodicity for small 2D and 3D systems. The case of a $2 \times 2$
square lattice is special, since no MC move connects the two possible
configurations. But for all other $L_x \times L_y$ lattices with $2
\le L_x,L_y \le 6$, we have checked that the number of different
configurations generated by the MC move, starting from an initial
reference configuration, coincides precisely with the exact
enumeration results of \cite{HamChains}. In particular, precisely
$17\,570\,172$ configurations are found on the $6 \times 6$ lattice.
We have similarly checked ergodicity on $2 \times 2 \times 2$, and $2
\times 2 \times 3$, and $2 \times 3 \times 3$ cubic lattices.

Assuming that ergodicity holds in general (except for the trivial $2
\times 2$ square lattice), we conclude that the MC process converges
towards the equilibrium distribution, i.e., that all HC configurations
are sampled with uniform probability.

\section{Properties}

In the case where the point ${\bf y}$ is an internal point on the same
chain as the end point ${\bf x}$, the identification of the correct
edge $({\bf y}{\bf z})$ to be removed necessitates tracing out the
loop formed when adding ${\bf x}{\bf y}$. So at worst, one MC move may
require a time $\sim N$. Since each move changes at most two links, it
is appropriate to evaluate the autocorrelation time $\tau$ (defined in
terms of the link overlap with the initial configuration) in units of
the number of MC moves per site. Just as in the $M=1$ case
\cite{Backbite} we find that in those units $\tau \sim L^z$ is of
order unity, independently of the system size. We conclude that the
dynamical exponent $z \approx 0$, proving our claim that the algorithm
is efficient.

Note that our algorithm provides unbiased sampling of configurations
of two chains of lengths $N_1$ and $N_2$, where only $N=N_1+N_2$ is
fixed. To pursue our goal---the study whether two chains of equal
length mix or not---we must impose the constraint $N_1=N_2$ in an
efficient way. Trial runs for $L=16$ and aspect ratios $r=1,2$ show
that the probability distribution $p(\rho)$ of the length ratio
$\rho=N_1/N$ is very close to being uniform: we have in fact
$|p(\rho)-1| < 0.01$ for all $\rho \in [0.02,0.98]$. It follows that
a configuration with $\rho=\frac12$ will occur on average once for
each unit of $\tau$.

It is thus tempting to collect a data point whenever $\rho=\frac12$.
There is however an important caveat. Suppose that the end points of
chain 1 are deeply immersed in chain 1, and same for chain 2, and that
the two chains happen to be of the same length. Then all MC moves will
be such that an end point ${\bf x}$ of some chain attacks a point
${\bf y}$ on the {\em same} chain, and never on the other, thus
conserving the length of either chain.  If data were collected
whenever $\rho=\frac12$, a lot of measurements would be made which
were spaced by just one MC move, and not one unit of $\tau$.  This
would strongly bias the results.  A better measuring protocol is to
collect a data point when $\rho=\frac12$ {\em and} require two data
points to be spaced by a least one autocorrelation time. With this
protocol, each data point gives an independent configuration with
$\rho=\frac12$, taken uniformly within the ensemble of chains of equal
length.

\begin{figure}
\includegraphics[width=8cm,angle=0]{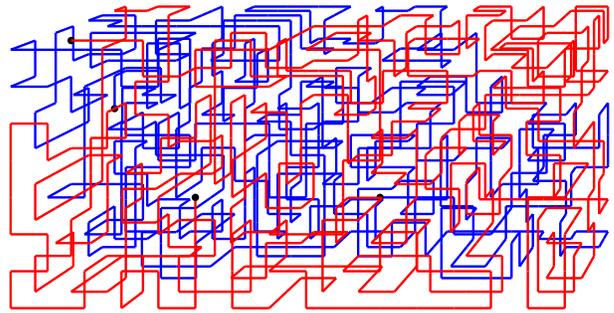}
\caption{(Color online.) A randomly chosen configuration of two chains of equal length on
 an $8 \times 8 \times 16$ cubic lattice.}
\label{fig:config}
\end{figure}

To avoid transient effects due to the choice of the initial state, we
discard in each run the first 100 MC moves per site, before starting
the data collection.  Fig.~\ref{fig:config} shows a measurement on a
rather small system obtained using this protocol. To the eye it
certainly looks like the two chains interpenetrate, but obtaining a
valid conclusion will require analysing the system more carefully.

\section{Scaling theory}

According to the standard scaling theory of polymers \cite{deGennes},
the number of HW of length $N$ is $\sim \mu^N N^{\gamma-1}$ for $N \gg
1$. Here $\mu$ is a lattice-dependent connective constant, and
$\gamma$ a universal conformational critical exponent. Alternatively,
$\gamma$ is determined by $P(x) \sim x^{2\gamma}$, where $P(x)$ is the
probability distribution of the end-to-end distance in the asymptotic
regime $x \ll 1$. Using this, the conformational exponent for HW has
been determined numerically as $\gamma = 0.94 \pm 0.02$
\cite{Backbite}. More recently, the consistent value $\gamma = 0.95
\pm 0.02$ has been reported \cite{Bohn}.

The value of $\gamma$ can be used in various ways to argue whether
$M=2$ chains will phase separate or not. One simple argument goes as
follows.  Suppose the two chains live in a volume $V$. In the demixing
scenario, each chain would occupy a volume $V/2$, and the number of
configurations would scale like that of two independent HW,
viz.~$\left[ \mu^{V/2} (V/2)^{\gamma-1} \right]^2 \sim \mu^V
V^{2(\gamma-1)}$. If, on the other hand, the two chains mixed, one
could make the assumption (A) that one end point of either chain
would be `close' and hence could be connected so as to form a single
HW in volume $V$. The number of configurations would then be $\mu^V
V^{\gamma-1}$. Comparing these, we conclude that entropy should drive
the chains to mix if and only if $\gamma < 1$. The above numerical
determinations of $\gamma$ are then in favour of the mixing scenario.

The applicability of the above argument clearly hinges on whether the
assumption (A) can be considered correct. Some partial justification
can be found in \cite{HamChains} in which the interactions of a single
HW with the surface of the system was carefully analyzed. In
particular, it was found that the end points possess a clear tendency
to avoid lattice sites of high coordination number, i.e., they are
attracted towards the surface. Since there are fewer surface sites
than bulk sites, this implies an enhanced probability for end points
to be close than would have been the case if they were uniformly
distributed throughout the volume. Moreover, the fact that $\gamma <
1$ means that end points attract one another entropically also in the
bulk.

But despite of such arguments, the assumption (A) is certainly a
rather weak spot in the above argument: the existence of an attraction
between end points does of course not mean that they are separated by
a single lattice spacing. So the above reasoning cannot be considered
as conclusive. The same is true for other variant scaling arguments
that we have tried out. We therefore turn to our numerical results to
determine whether the two chains will mix.

\section{Numerics} 

We have used our MC algorithm to generate an extensive number of
independent configurations for the two-chain problem on $L \times L
\times (rL)$ cubic lattices with $L=16$ or $L=20$ and various aspect ratios $1
\le r \le 10$. In the largest simulation ($L=20$ and $r=10$) each of
the chains had length $\frac{r}{2} L^3 = 40\,000$ monomers and up to
$1.1 \cdot 10^5$ statistically independent configurations were generated.
The boundary conditions are non-periodic (hard wall confinement) in all
three lattice directions, as shown in Fig.~\ref{fig:config}.

\begin{table}
\begin{tabular}{l|l|l}
  $r$ & $\langle \sigma_z \rangle$ & $w_\star$ \\ \hline
  $1$ & $0.28231 \pm 0.00005$ & $0.3632 \pm 0.0005$ \\
  $2$ & $0.27675 \pm 0.00008$ & $0.3580 \pm 0.0005$ \\
  $3$ & $0.27094 \pm 0.00011$ & $0.3560 \pm 0.0005$ \\
  $4$ & $0.26527 \pm 0.00019$ & $0.3550 \pm 0.0003$ \\
  $5$ & $0.26002 \pm 0.00015$ & $0.3546 \pm 0.0003$ \\
\end{tabular}
\caption{\label{tab:stddev}
  For various aspect ratios $r$, we show:
  (1) the averaged standard deviations $\langle \sigma_z \rangle$ of the
  $z$-coordinate of a chain in the fully-packed two-chain problem,
  and (2) the monomer fugacity $w_\star$ that will make a single
  chain fill out one half of the available volume.}
\end{table}

In a first series of runs, we produced ${\cal N}_1 \simeq 3 \cdot
10^4$ configurations for each $r$. For each of these, the standard
deviation $\sigma_z = \sqrt{\langle z^2 \rangle - \langle z
  \rangle^2}$ of the $z$-coordinates of the monomers---rescaled so
that $z \in [0,1]$---was computed separately for each chain, and an
average value $\langle \sigma_z \rangle$ was extracted from the $2
{\cal N}_1$ measures. The results for $L=16$ systems are given in the
left part of Table~\ref{tab:stddev}.  They can be compared to the
result $\sigma_z^0 = \frac{1}{2 \sqrt{3}} \simeq 0.28868$ which would
be obtained in the case of full mixing, where the mass of each chain
is uniformly distributed in the $z$-direction.

The results for $\langle \sigma_z \rangle$ indicate that the two
chains tend to demix, and that this tendency grows with increasing
$r$.  Of course, one is far away from the value $\frac12 \sigma_z^0$
which would result in the hypothetical situation of complete demixing,
where one chain goes completely to the left of the box, and the other
to the right. Note also that the difference from $\sigma_z^0$ is
clearly measurable---albeit tiny---for $r=1$ where the three
coordinates are equivalent.

\begin{figure}
\includegraphics[width=8cm,angle=0]{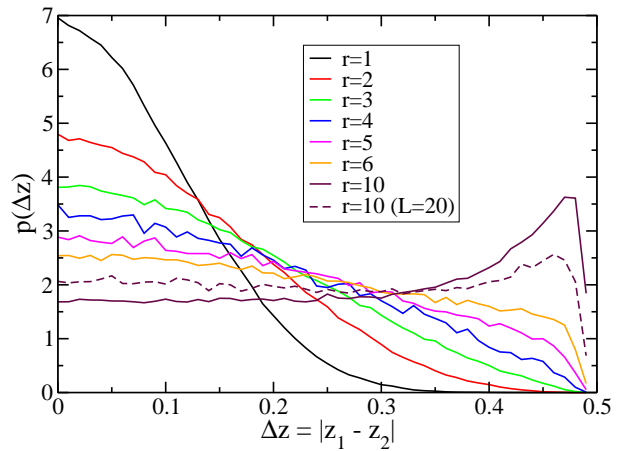}
\caption{(Color online.) Probability distribution $p(\Delta z)$ of the difference of
  center of mass coordinates $\Delta z = |z_1 - z_2|$ of the two
  chains, for $L=16$ lattices with various aspect ratios $r$. The
  dashed curve is for $L=20$ and $r=10$.}
\label{fig:dist}
\end{figure}

To give more evidence for the demixing, and characterise it more
precisely, we turn to a second series of runs. In this series we
construct the probability density $p(\Delta z)$ of the quantity
$\Delta z = |z_1 - z_2|$, where $z_i = \langle z \rangle$ is the
$z$-coordinate of the center of mass of the monomers in chain $i$ (for
$i=1,2$). Obviously, constructing a whole probability distribution
calls for better statistics, so in this series we produced ${\cal N}_2
\simeq 1.1 \cdot 10^5$ configurations for each $r$.  The results for
$p(\Delta z)$ are shown in Fig.~\ref{fig:dist}.  To eliminate
fluctuations arising from the lattice discretization, the values of
$\Delta z$ have been arranged into 50 `running' bins; the statistical
error bars can be judged from the remaining small ripples on the
curves.

\begin{figure}
\includegraphics[width=8.5cm,angle=0]{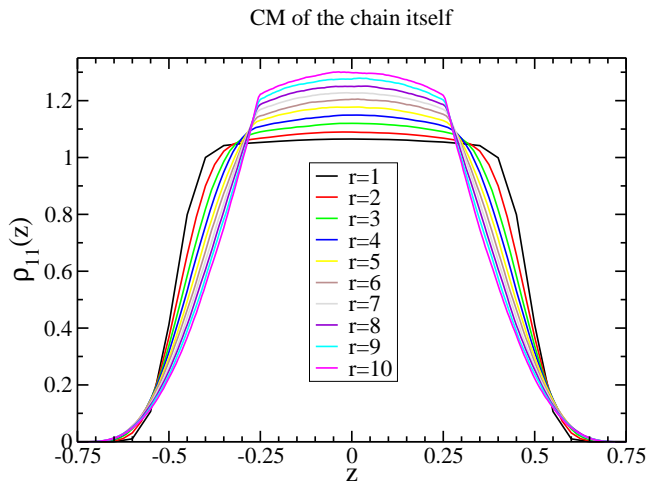}
\caption{(Color online.) Mass density $\rho_{11}(z)$ of one of the chains, measured
  relative to its center of mass coordinate $z_1$. The data are for
  $L=20$ lattices with various aspect ratios $r$.}
\label{fig:same}
\end{figure}

\begin{figure}
\includegraphics[width=8.5cm,angle=0]{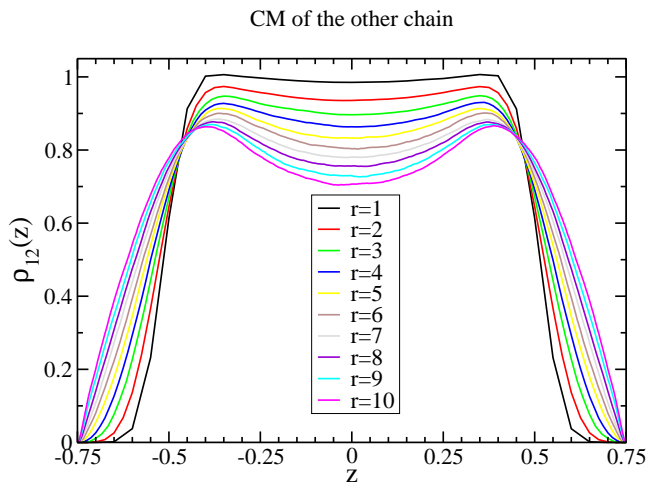}
\caption{(Color online.) Mass density $\rho_{12}(z)$ of chain 1, measured
  relative to the center of mass coordinate $z_2$ of the other chain.
  The data are for
  $L=20$ lattices with various aspect ratios $r$.}
\label{fig:other}
\end{figure}

In a third series of runs we have focused on the mass density
$\rho_{ij}(z)$ of chain $i$, measured relative to the center of mass $z_j$ of
chain $j$ (for $i=1,2$). The normalization is chosen so that $\int
\rho_{ij}(z) \, {\rm d}z = 1$. Demixing can be detected via a
different behavior for the density of a chain in its own center of
mass ($\rho_{11} = \rho_{22}$ by symmetry) and for the density of a
chain in the center of mass of the other chain ($\rho_{12} =
\rho_{21}$).  For these measurements we have taken $L=20$ and produced
up to ${\cal N}_3 \simeq 3 \cdot 10^4$ configurations for each $r$.
Instead of using bins, we have simply rounded the center of mass
coordinates as expressed in lattice spacings (i.e., $r L z_j$) to the
nearest integer.  The results for $\rho_{11}(z)$ are given in
Fig.~\ref{fig:same}, and those for $\rho_{12}(z)$ in
Fig.~\ref{fig:other}. Both functions should of course be symmetric
upon reversing the coordinate axis, $\rho_{ij}(z) = \rho_{ij}(-z)$,
and the error bars can be judged from the slight deviations from this
symmetry. It is seen from the figures that when $r$ increases,
$\rho_{11}(z)$ tends to become a more narrow distribution around the
origin, whereas $\rho_{12}(z)$ spreads out and develops a depletion
near the origin.  This is again clear evidence of demixing.

\section{Discussion} 

The data in Fig.~\ref{fig:dist} give clear evidence for demixing as
$r$ increases. For small $r$ we see a rather broad distribution
$p(\Delta z)$, localized around the origin. At $r \simeq 7$, the
distribution becomes almost uniform, and for higher $r$ a peak near
$\Delta z = \frac12$, the maximal possible value, develops on a
close-to-uniform background. To make sure that this conclusion is not
a finite-size artefact, we recomputed the $r=10$ curve for a larger
size $L=20$. Note that this computation required several years of CPU
time. The result, shown as a dashed curve in Fig.~\ref{fig:dist}, does
not deviate substantially from the $L=16$ result.

\begin{figure}
\includegraphics[width=8.5cm,angle=0]{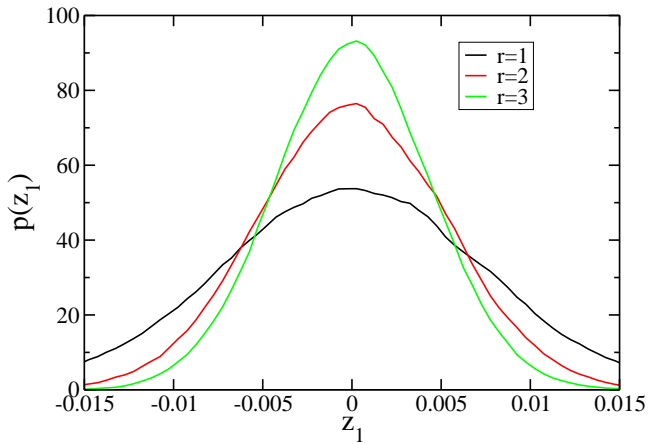}
\caption{(Color online.) Probability distribution $p(z_1)$ of the centre of mass
  coordinate $z_1$ of a single chain, constrained to occupy
  precisely one half of the box, for different aspect ratios $r$.}
\label{fig:onechain}
\end{figure}

It is useful to compare this situation to that of a single non-fully
packed chain that is constrained to take up precisely one half of the
available volume. To study this, we employ the non-fully packed
version of the $M=1$ algorithm, discussed very briefly at the end of
\cite{Backbite}. To set the peak of the monomer concentration $\rho =
N/V$ precisely at $\rho = \frac{1}{2}$, one needs to tune the
Boltzmann weight $w$ of a monomer. Still for $L=16$, we find that this
is obtained for $w=w_\star$, where values of $w_\star$ for different
$r$ are given in the right part of Table~\ref{tab:stddev}. This allows
us to take a series of data for which each configuration has $\rho =
\frac12$ precisely. Based on a set of ${\cal N}_3 \simeq 10^4$ data
points, we then construct the probability distribution $p(z_1)$ of the
center of mass $z_1$ of the chain. For convenience, we here subtract
$\frac12$ from the $z$-coordinates so as to set the origin in the
middle of the box; the support of $p(z_1)$ is then $z_1 \in
[-\frac14,\frac14]$. The results for $p(z_1)$, using this time 200
bins, are given in Fig.~\ref{fig:onechain}.

It is of course completely predictable that these distributions are
Gaussian; we find indeed $p(z_1) \simeq a \exp \big(-b (z_1)^2\big)$
with $b = (0.91 \pm 0.01) \times r$. But more importantly, comparing
Figs.~\ref{fig:dist}--\ref{fig:onechain}, one can appreciate that even
for $r=1$ the distribution $p(\Delta z)$ is indeed very broad.  This
corroborates our previous claims that even for $r=1$ the two-chain
problem indeed exhibits demixing.

\begin{figure}
\includegraphics[width=8.5cm,angle=0]{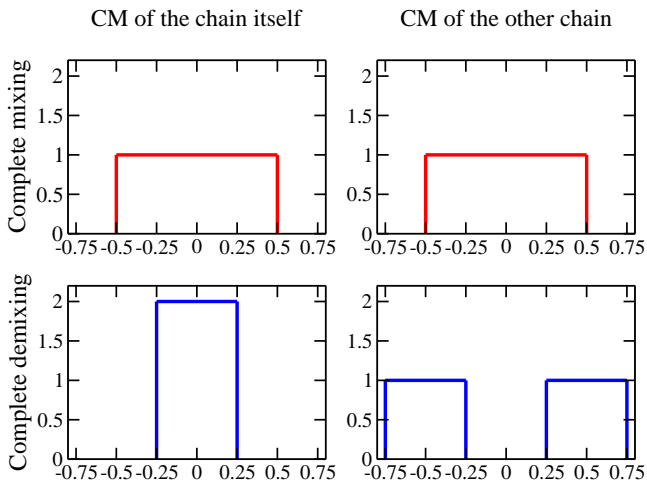}
\caption{(Color online.) The appearance of the distributions $\rho_{11}(z)$
  (left panels, cf.\ Fig.~\ref{fig:same}) and $\rho_{12}(z)$
  (right panels, cf.\ Fig.~\ref{fig:other}) in the two extremal
  and hypothetical situations of complete mixing (top panels, red
  or light gray curves) and complete demixing (bottom panels, blue or dark gray curves).}
\label{fig:complete}
\end{figure}

It is useful to compare the data in
Figs.~\ref{fig:same}--\ref{fig:other} to the hypothetical situations
of complete mixing (with both chains uniformly distributed in the
$z$-direction) and of complete demixing (with one chain occupying the
left half of the box, $0 < z < \frac12$, and the other chain the right
half, $\frac12 < z < 1$). In these two extremal situations, the mass
distributions $\rho_{11}(z)$ and $\rho_{12}(z)$ would be those shown
in Fig.~\ref{fig:complete}. The data of
Figs.~\ref{fig:same}--\ref{fig:other} bear some resemblance to a
crossover (with increasing $r$) between these extremes. A closer
examination of the data however indicates that for $r \gg 1$ the
distributions $\rho_{ij}(z)$ will {\em not} coincide with those of the
complete demixing scenario. (For instance, the tails in
Fig.~\ref{fig:same} extending out to $z = \pm \frac34$ seem to be
robust upon increasing $r$.)  Note also that in terms of the
distribution $p(\Delta z)$ of Fig.~\ref{fig:dist} complete demixing
would correspond to a Dirac delta function $\delta(z-\frac12)$,
whereas the actual data are indicative of a close-to-uniform
distribution on $[0,\frac12]$, plus possible a finite peak near
$z=\frac12$. In other words, the numerics strongly supports the idea
that the physics in the limit $\lim_{r \to \infty} \lim_{L \to
  \infty}$ (with the thermodynamic limit being taken first) is
non-trivial and corresponds to some kind of ``partial demixing''.

\section{Conclusion} 

We have studied a problem of two compact polymer chains of equal
length in a three-dimensional box of various aspect ratios $r$. We
have devised an efficient Monte Carlo method for sampling all
configurations of this problem with uniform probability.  Naive
scaling theory, combined with existing evaluations of the
conformational exponent $\gamma$ for a Hamiltonian walk, indicates
that the two chains should mix completely. This same conclusion would
be suggested by the Flory theorem \cite{Flory} for a polymer melt of
many chains, or from a blob-type analysis for a simple piston model of
just two chains \cite{Jun}. However, our numerical simulations
clearly show that the two chains do in fact phase separate, even for
$r=1$, and that this demixing becomes more pronounced, albeit not perfect,
in the limit of large $r$.

Clearly, further work is needed to see whether this conclusion can be
reconciled with some improved scaling theory, specifically adapted to
a melt consisting of just a few chains. The demixing scenario obtained
in this paper is somewhat subtle, since it is not due to the additive
terms of the free energy. Further study is therefore required to
assess whether the effect is stable towards introducing more
parameters into the model (non-identical chain lengths, specific
chemical interactions at the chain ends, off-lattice modelling, etc.).
It would also be interesting to see whether demixing can be realized
experimentally.

\subsection*{Acknowledgments}

The author thanks J.~Kondev, H.~Orland and
P.~Wiggins for stimulating discussions.  This work was supported by
the Agence Nationale de la Recherche (grant ANR-06-BLAN-0124-03).

\end{document}